\documentclass[aps,pre,preprint,superscriptaddress]{revtex4-2}
\usepackage{graphicx}
\usepackage{bm}
\usepackage{subcaption}
\usepackage{amsmath,amssymb}
\usepackage[T1]{fontenc}

\begin{document}

\title{Solving Lyapunov equations for electrically driven ternary electrolytes - application to long-range van der Waals interactions}
\date{\today}
\author{Guangle Du}
\affiliation{ Center of Materials Science and Optoelectronics Engineering, College of Materials Science and Opto-Electronic Technology, University of Chinese Academy of Sciences (UCAS), Beijing 100049, China}
\author{Bing Miao}\email{bmiao@ucas.ac.cn}
\affiliation{ Center of Materials Science and Optoelectronics Engineering, College of Materials Science and Opto-Electronic Technology, University of Chinese Academy of Sciences (UCAS), Beijing 100049, China}
\author{David S. Dean}\email{david.dean@u-bordeaux.fr}
\affiliation{Universit\'e Bordeaux, CNRS, LOMA, UMR 5798, F-33400 Talence, France}
\affiliation{ Kavli Institute for Theoretical Sciences, University of Chinese Academy of Sciences (UCAS), Beijing 100049, China}
\begin{abstract}
Stochastic density functional theory (SDFT) has been widely used to study the out of equilibrium properties of electrolyte solutions. Examples include investigations of electrical conductivity---both within and beyond linear response---and modifications of thermal van der Waals interactions in driven electrolytes. Within the approximation scheme derived from linearizing SDFT for fluctuations around mean densities, the steady state correlation functions between the $N$ ionic species are governed by linear Lyapunov equations of degree $N(N+1)/2$. Consequently, the system's complexity increases significantly when transitioning from binary to ternary electrolytes, and few analytical results exist for the latter.
In this paper, we demonstrate how---for the specific case of electrolytes---the Lyapunov equations can be reduced to a system of $N$ linear equations. We apply this reduction to compute the long-range component of the van der Waals interaction between two slabs containing a ternary electrolyte under an applied electric field parallel to the slabs. Unlike the binary electrolyte case, we show that the resulting van der Waals interaction for a ternary electrolyte depends on the ionic species' diffusion coefficients, highlighting its inherently out of equilibrium nature. 
\end{abstract}

\maketitle

\section{Introduction}
\label{intro}
In recent years stochastic density functional theory (SDFT) \cite{kawasakiStochasticModelSlow1994,deanLangevinEquationDensity1996} has been applied to study the out of equilibrium behavior of electrolytes driven by external electric fields \cite{demeryConductivityStrongElectrolytes2016,avniConductivityConcentratedElectrolytes2022,andelmanwien22,bernardAnalyticalTheoriesConductivity2023,andelmanac24,vincentwien24,illienStochasticDensityFunctional2024}. These studies started by generalizing Onsager's results  \cite{onsagerWienEffectSimple1957} on electrolyte conductivity to arbitrary dimensions and for additional (non-electrostatic) interactions for a purely Brownian model of electrolytes. These results were subsequently extended to incorporate hydrodynamic interactions and compute AC conductivities. The starting point in all of these computations is the formulation of the Brownian dynamics of the ionic species in terms of the evolution of their density fields \cite{kawasakiStochasticModelSlow1994,deanLangevinEquationDensity1996}. These stochastic density functional equations are analytically intractable due to their non-linearity and the presence of multiplicative noise. However, if the equations are expanded about the mean density of each ionic species in terms of their fluctuations and are then linearized in terms of these density fluctuations, the resulting equations describe Gaussian fluctuations and belong to the model B (conserved) form of phase ordering kinetics \cite{brayTheoryPhaseorderingKinetics2002}. The resulting dynamical approximation then turns out to be an extension of the Debye-H\"uckel approximation to out of equilibrium systems. In this approximation scheme one can derive expressions for the correlation functions of the $N$ density fluctuation fields $n_{i}({\bf r})$ (where the index $i$ indicates the species type)
\begin{equation}
C_{ij}({\bf r},{\bf r}') = \langle n_{i}({\bf r})n_{j}({\bf r}')\rangle \label{cfs}
\end{equation}
in the nonequilibrium steady state where the electrolyte is driven by an applied constant external electric field. From this correlation function one can compute several objects of interest for electrolytes driven by electric fields, as mentioned above these include the electrical conductivity and long-range thermal van der Waals forces, but also one can compute the viscosity of electrolyte solutions \cite{robin23}.

\begin{figure}[!t]
  \centering
  \includegraphics[width=0.9\linewidth]{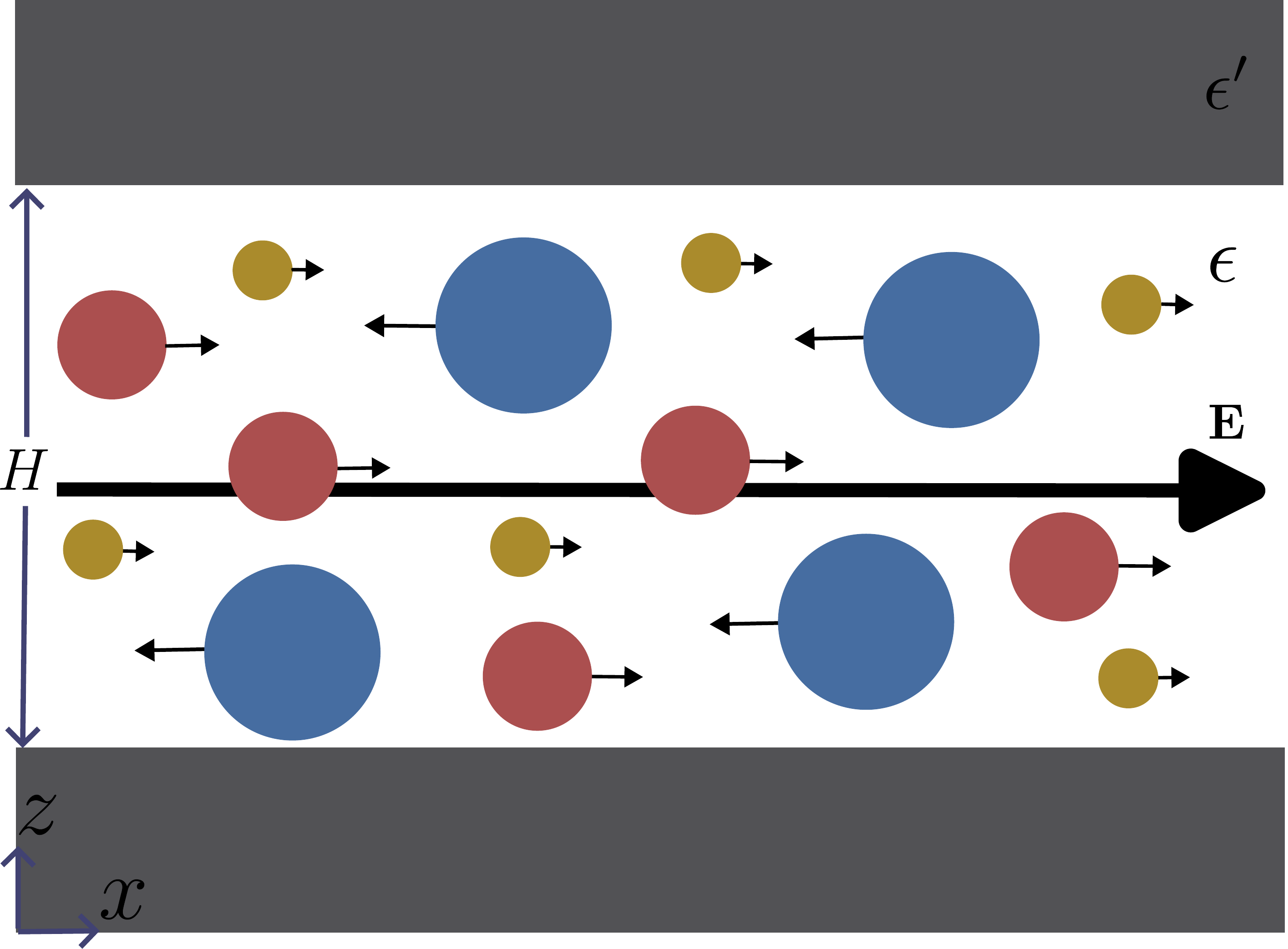}
  \caption{\label{fig:schematic} Schematic of three species electrolyte driven by an external electric field.}
\end{figure}

The equations obeyed by the correlation functions in Eq. (\ref{cfs}) are known as Lyapunov equations and occur generically for Gaussian fluctuations out of equilibrium \cite{zwanzigNonequilibriumStatisticalMechanics2001}. These Lyapunov equations  take a matrix form and, due to the symmetry of the correlation functions, are composed of  $N(N+1)/2$ equations corresponding to the number of independent terms of the correlation functions $C_{ij}({\bf r},{\bf r}')$. In Fourier space, the Lyapunov equations decouple in terms of the individual Fourier modes but  still remain  rather complex. They  yield complicated and rich  expressions, even  for simple binary ($N=2$) electrolytes. The extension from binary to ternary electrolytes represents an increase from a system of $3$ equations to $6$ equations, meaning that the matrix inversion required to solve the system is considerably more complicated. The aim of this paper is to show how the systems of $N(N+1)/2$ linear equations arising for $N$ component electrolytes can be reduced to a system of $N$ equations for the particular case of electrolytes.

We hope that this formulation will be useful for future studies of multicomponent electrolytes within the SDFT formalism. As a first  application,  we consider here the case of the long-range thermal van der Waals interaction between parallel semi-infinite poor dielectric (that is to say having very low dielectric constants compared  to the electrolyte solvent  water) interfaces separated by an electrolyte solution which is driven by an electric field of magnitude $E$ parallel to the interfaces - see Fig. (\ref{fig:schematic}). This problem has been recently studied \cite{mahdisoltaniLongRangeFluctuationInducedForces2021,duCorrelationDecouplingCasimir2024,Du2025b} and is an example of how what are generically called fluctuation-induced forces \cite{lifshitzTheoryMolecularAttractive1956,woodsMaterialsPerspectiveCasimir2016,klimchitskayaCasimirForceReal2009a,woodsPerspectiveRecentFuture2020a,dantchevCriticalCasimirEffect2023}, including Casimir, van der Waals forces, critical or thermal Casimir forces, behave in nonequilibrium settings. Here the system is out of equilibrium due to the driving which generates a current. Other nonequilibrium situations due to temperature differences and quenches in both quantum \cite{antezzaCasimirLifshitzForceOut2006,antezzaCasimirLifshitzForceOut2008,antezzaNewAsymptoticBehavior2005a,bimonteScatteringApproachCasimir2009,dorofeyevForceAttractionTwo1998,krugerNonequilibriumElectromagneticFluctuations2011,luOutofequilibriumThermalCasimir2015,messinaCasimirLifshitzForceOut2011} and classical \cite{gambassiCriticalDynamicsThin2006,gambassiRelaxationPhenomenaCriticality2008,deanNonequilibriumBehaviorPseudoCasimir2009,deanOutofequilibriumBehaviorCasimirtype2010a,deanRelaxationThermalCasimir2014} systems have been studied.
In the absence of driving, the thermal component of the van der Waals interaction is well known and is given at large interface separations $H$ by the  screened interaction \cite{mahantyDispersionForces1976,parsegianVanWaalsForces2006,netzStaticVanWaals2001,jancoviciScreeningClassicalCasimir2004,deanElectrostaticFluctuationsSoap2002}
\begin{equation}
f_{\rm sc}(H)\simeq -\frac{k_{\text{B}} T \kappa^2}{4\pi  H} \exp(-2\kappa H),
\end{equation}
where $k_{\text{B}}$ is Boltzmann's constant and $T$ is the temperature. The term $\kappa$ is the inverse Debye screening length and is given by
\begin{equation}
\kappa^2 =  \sum_{i=1}^N\frac{\beta q_i^2 \bar{\rho}_i}{\epsilon},
\end{equation}
where $q_i$ is the charge of ionic species $i$ and $\bar{\rho}_i$ is its mean number density. In addition $\beta = 1/(k_{\text{B}} T)$ and
$\epsilon$ is the dielectric constant of the solvent. In all the cases we will consider we impose the global constraint of electroneutrality
\begin{equation}
\sum_{i=1}^N \bar{\rho}_i q_i=0. \label{eneut}
\end{equation}

In the above we have assumed that we are in the limit where the dielectric constant of the dielectric interfaces $\epsilon'$ is much smaller than the dielectric constant of the solvent $\epsilon$, i.e., $\epsilon' \ll \epsilon$. The above result is classic and well established and can be derived in a variety of different manners in the limit where the Debye-H\"uckel approximation is valid. This screened result should be compared with the long-range thermal van der Waals interaction in the case where there is no electrolyte and just the dielectric solvent. In this particular dielectric configuration the thermal van der Waals interaction is given by the universal thermal Casimir form \cite{parsegianVanWaalsForces2006}
\begin{equation}
f_{\rm vdW}(H)= -\frac{k_{\mathrm{B}}T\zeta(3)}{8\pi H^3},\label{vdW}
\end{equation}
where $\zeta(z)$ denotes the Riemann zeta function. In equilibrium we therefore see that the presence of electrolyte has a very strong influence, effectively killing off the long-range thermal component of the van der Waals interaction between the two slabs.

In Ref.~\cite{Du2025b}, the results of Ref.~\cite{duCorrelationDecouplingCasimir2024} were extended to the case of general binary but non-symmetric electrolytes. By this, we mean electrolytes where the mean ionic densities $\bar{\rho}_i$, charges $q_i$ and diffusion constants $D_i$ of each species $i$ are different, while the electroneutrality condition Eq. (\ref{eneut}) holds. The results of Ref.~\cite{Du2025b} showed that for a binary electrolyte an applied electric field induces a long-range component to the thermal van der Waals contribution given by
\begin{align}
f^{\text{lr}}_{\text{t}}(H) =
\frac{\zeta(3)}{8\pi \beta H^3} \left[ 1 - \frac{1}{\sqrt{1+ \frac{\beta \epsilon E^2}{\bar{\rho}_1 + \bar{\rho}_2}}} \right].\label{2elec}
\end{align}
This result is remarkable as it shows that driving by an external electric field parallel to the plates induces a long-range interaction and effectively disrupts the screening mechanism. Furthermore, another interesting feature of this result is that the long-range force is independent of the diffusion constants $D_i$ of the ionic species. We note that this result is in contrast with the problem studied in Ref.~\cite{deanNonequilibriumTuningThermal2016} where the interaction between two parallel plates containing mobile charges on a uniform neutralizing background was considered. Here, in equilibrium,  the interaction between the plates at long distance is not screened (as there are no intervening charges) and takes the universal thermal Casimir form
given in Eq. (\ref{vdW}). When an electric field is applied to the charges in one of the plates along the plate, a current flows and the long-range interaction between the two plates takes the form \cite{deanNonequilibriumTuningThermal2016}
\begin{equation}
f_{\rm vdW}(H)= -\frac{k_{\mathrm{B}}T {\cal H}}{8\pi H^3},\label{vdWE}
\end{equation}
where ${\cal H}$ is an effective constant that depends on the magnitude of the electric field, the density and charges of the mobile ions, but also the diffusion constants of the ions. However we see that Eq. (\ref{2elec}) only depends on static quantities, being independent of the ionic diffusion coefficients. This observation leads to the possibility that the driven state in the problem with intervening electrolyte has some hidden  equilibrium nature. The study of  Ref.~\cite{Du2025b}  also explored the long-range interaction for a driven ternary electrolyte, consisting  of two salts having either a common cation or anion (for example NaCl and KCl). However, due to their complexity,  the relevant Lyapunov equations could only be solved for systems where all the diffusion constants were assumed to be equal, even when algebraic computer packages such as {\em Mathematica} are employed. The resulting long-range force  in these special (and of course physically unrealistic) cases is again independent of the diffusion constants \cite{Du2025b},  this fact is however also obvious from dimensional analysis.

The purpose of this paper is  therefore to develop a method to reduce the dimension of the Lyapunov equations from a linear size of $N(N+1)/2$ to $N$. This allows us to, but still via computer algebra, solve the resulting Lyapunov equations in the low wave number limit necessary to compute the long-range van der Waals interaction. This solution shows that the resulting force {\em does} in general depend on the individual ionic diffusion constants for a ternary electrolyte and we give the resulting, rather complicated, expression for the force and explore how it depends on the ionic diffusivities.

The paper is organized as follows. In Sec.~\ref{sdft} we briefly recall the SDFT formalism and its  form in the approximation where only linear terms in the density fluctuations are taken into account.  We then show in Sec.~\ref{lyap} how the linearized SDFT can be rescaled in such a way that the resulting Lyapunov equations are reducible 
to a linear system of size $N$. In Sec.~\ref{vdw} we then employ this reduction to compute the long-range van der Waals interaction for two dielectric media separated by a ternary electrolyte solution which is driven by an electric field parallel to the surfaces of the two media. In Sec.~\ref{disc} we discuss the physical implications of the calculation
in general and for some specific electrolyte mixtures. In Sec.~\ref{conc} we give our conclusions.

\section{Stochastic density functional theory}
\label{sdft}
When an external electric field is applied, the system is driven out of equilibrium, and thus in order to study the system one must specify the dynamics. Here, we assume that the ions undergo Brownian motion in an aqueous solvent, and that forces acting on them arise from both the applied electric field and their mutual electrostatic interactions.
We adopt a SDFT in which the evolution of ionic densities is described by \cite{deanLangevinEquationDensity1996}
\begin{align}
\label{eqs:DK-nonlinear}
\partial_t \rho_i(\mathbf{x},t) = D_i \nabla^2 \rho_i + D_i \beta q_i \bm{\nabla} \cdot \left[\rho_i (\nabla \phi - \mathbf{E})\right]
 + \bm{\nabla} \cdot \sqrt{2D_i \rho_i} \, \bm{\eta}_i(\mathbf{x},t),
\end{align} 
where \(\bm{\eta}_i(\mathbf{x},t)\) are spatiotemporal Gaussian white noise vector fields
\begin{align}
\langle \eta_{i,\alpha}(\mathbf{x},t) \eta_{j,\beta}(\mathbf{x}',t')\rangle = \delta_{ij} \delta_{\alpha\beta} \delta(\mathbf{x}-\mathbf{x}') \delta(t-t'),
\end{align}
where \(i\) and \(j\) denote the species indices, and \(\alpha\) and \(\beta\) denote the vector component indices, respectively. The term 
$\phi$ denotes the electrostatic potential computed from Poisson's equation
\begin{align}
\bm{\nabla}\cdot \epsilon \bm{\nabla} \phi = - \sum_{j=1}^N q_j \rho_j.
\end{align}

Eq.~\eqref{eqs:DK-nonlinear} is nonlinear and hard to solve, but becomes tractable after linearization
via expanding the ionic densities \(\rho_i = \bar{\rho}_i + n_i\), where \(\bar{\rho}_i\) is constant due to the assumption of no charge on the dielectric interfaces
\begin{align} \label{eqs:DK-linear}
\partial_t n_i =&\, D_i \left( \nabla^2 n_i - \frac{\beta q_i \bar{\rho}_i}{\epsilon} \sum_{j=1}^N q_j n_j - \beta q_i \mathbf{E} \cdot \bm{\nabla} n_i \right) \nonumber\\
& +  \bm{\nabla} \cdot \sqrt{2D_i \bar{\rho}_i} \, \bm{\eta}_i.
\end{align}

The presence of the dielectric boundaries in the problem means that Eq.~\eqref{eqs:DK-linear} must be supplemented with the no flux boundary conditions for  deterministic part of the current
\begin{align} \label{eq:bc}
\partial_z n_i + \beta q_i \bar{\rho}_i \partial_z \phi = 0,
\end{align}
and the random part of the current
\begin{equation}
{\eta}_{iz}(\mathbf{x},t)=0, \label{f2}
\end{equation}
at each dielectric interface.
In the derivation of the SDFT \cite{deanLangevinEquationDensity1996} the noise \(\bm{\eta}_{i}(\mathbf{x},t)\)  has  the Ito interpretation and is independent of the state of the density field. As such, the  deterministic part of current in  Eq.~\eqref{eqs:DK-linear} (the first three terms) and the random part (the last term)  must be independent. This independence means that   both currents must have no flux boundary conditions and not just the total sum.

For the limit \(\epsilon \gg \epsilon'\),
it is easy to see from the full no flux boundary condition in Eq.~\eqref{eq:bc} that  boundary conditions for \(n_i\) are  Neumann, because, crucially, our choice of dielectrics means that those for  \(\phi\) are also Neumann.
This means that we can use  the following Fourier expansion
\begin{align}
n_i(\mathbf{x}_{\|}, z, t) =&\, \int \frac{\mathrm{d}\mathbf{k}}{(2\pi)^2} \sum_{n=0}^{\infty} \frac{1}{\sqrt{N_n}} \tilde{n}_{i,n}(\mathbf{k}, t) \nonumber\\
& \times \exp(\mathrm{i} \mathbf{k}\cdot \mathbf{x}_{\|}) \cos (p_n z),
\end{align}
where \(N_0 = H\) and \(N_n = H/2\) for \(n\ge 1\), \(p_n = n\pi/H\) enforces the Neumann boundary conditions. In Fourier space, we obtain
\begin{align}
\partial_t \tilde{n}_{i,n} =&\, - D_i(k^2 + p_n^2 + \beta q_i E \mathrm{i} k_x) \tilde{n}_{i,n} \nonumber\\
&\, - D_i \frac{\beta q_i \bar{\rho}_i}{\epsilon} \sum_{j=1}^N q_j \tilde{n}_{j,n} + \xi_{i,n}(\mathbf{k},t),
\end{align}
where the noise term has correlation function
\begin{equation}
\langle \xi_{i,n}(\mathbf{k}, t) \xi_{j,m}(\mathbf{k}', t')\rangle = (2\pi)^2 \delta(t-t') \delta(\mathbf{k}+\mathbf{k}') \delta_{ij} \delta_{nm} 2 D_i \bar{\rho}_i (k^2 + p_n^2).
\end{equation}

The equal-time correlation functions in the steady state are denoted by
\begin{align}
\langle \tilde{n}_{i,n} (\mathbf{k}) \tilde{n}_{j,n}(\mathbf{k}')\rangle = (2\pi)^2 \delta(\mathbf{k}+\mathbf{k}') \tilde{C}_{ij,n}(\mathbf{k}),
\end{align}
where the correlation matrix \(\tilde{C}_{ij,n}(\mathbf{k})\) obeys the Lyapunov equation \cite{zwanzigNonequilibriumStatisticalMechanics2001}
\begin{align}
\label{eq:Lyapunov-before}
A \tilde{C}_n + \tilde{C}_n A^{\dagger} = 2R,
\end{align}
with
\begin{subequations}
\begin{align}
A_{ij} &= D_i \left[  (k^2 + p_n^2 + \beta q_i E \mathrm{i} k_x) \delta_{ij} + \frac{\beta q_i \bar{\rho}_i}{\epsilon} q_j \right], \\
R_{ij} &= (k^2 + p_n^2) D_i \bar{\rho}_i \delta_{ij}.
\end{align}
\end{subequations}
We note that in a bulk system, in order to study conductivity or viscosity for instance, the equations take the same form but where $k^2 + p_n^2$ is replaced by $k^2$, the magnitude of the three-dimensional wave vector. 
\section{Rescaled Lyapunov equation}
\label{lyap}
To facilitate solving the Lyapunov equation, we observe that the coefficient matrix \(A\) consists of two parts: a diagonal part and a outer product part.
The outer product part can be made symmetric by rescaling fluctuating density field and the spatiotemporal noise field
\begin{align}
\tilde{m}_{i,n} = \frac{\tilde{n}_{i,n}}{\sqrt{D_i \bar{\rho}_i}}, \quad \Xi_{i,n} = \frac{\xi_{i,n}}{\sqrt{D_i \bar{\rho}_i}}.
\end{align}
We now define \(\kappa_i = \sqrt{\beta \bar{\rho}_i/\epsilon}\, q_i\), \(K_i = \sqrt{D_i}\, \kappa_i\) and \(v_i = D_i \beta q_i E\) is the bare velocity of an ion of type $i$ in the external field.
The rescaled SDFT  equation is
\begin{align}
\partial_t \tilde{m}_{i,n} = - [D_i(k^2 + p_n^2) + v_i \mathrm{i} k_x)] \tilde{m}_{i,n} - \sum_{j=1}^N K_i K_j \tilde{m}_{j,n} + \Xi_{i,n},
\end{align}
where
\begin{equation}
\langle \Xi_{i,n}(\mathbf{k}, t) \Xi_{j,m}(\mathbf{k}', t')\rangle = (2\pi)^2 \delta(t-t') \delta(\mathbf{k}+\mathbf{k}') \delta_{ij} \delta_{nm} 2 (k^2 + p_n^2).
\end{equation}

The equal-time correlation functions of the rescaled fluctuating density field in the steady state is
\begin{align}
\langle \tilde{n}_{i,n} (\mathbf{k}) \tilde{n}_{j,n}(\mathbf{k}')\rangle = (2\pi)^2 \delta(\mathbf{k}+\mathbf{k}') \tilde{C}'_{ij,n}(\mathbf{k}),
\end{align}
where the correlation matrix \(\tilde{C}'_{ij,n}(\mathbf{k})\) obeys the modfied Lyapunov equation
\begin{align}
A' C' + C' A'^{\dagger} = 2R',
\end{align}
where
\begin{align}
A'_{ij} = \delta_{ij} [D_i (k^2+p_n^2) + \mathrm{i} k_x v_i] + K_i K_j, \quad R'_{ij} = \delta_{ij} (k^2+p_n^2).
\end{align}
In component form, the Lyapunov equation is
\begin{align}
[(D_i + D_j)(k^2+p_n^2) + \mathrm{i} k_x (v_i - v_j)] C'_{ij} + \sum_{k=1}^N \left(K_i K_k C'_{kj} + C'_{ik} K_k K_j\right) = 2 (k^2+p_n^2) \delta_{ij}.\label{rslyap}
\end{align}
The original correlation matrix (before rescaling) is then recovered as
\begin{align}
\tilde{C}_{ij,n} = \sqrt{D_i \bar{\rho}_i D_j \bar{\rho}_j} \, \tilde{C}'_{ij,n}.
\end{align}

We now show the key point of our paper: how we can  reduce
the rescaled linear Lyapunov equation, Eq.~\eqref{rslyap}, of degree \(N(N+1)/2\) to a system of \(N\) linear equations.

By definition, \(\tilde{C}'\) is necessarily an Hermitian matrix, i.e., \(\tilde{C}'_{ij}\) = \(\tilde{C}'^{*}_{ji}\).
We start by defining the vector
\begin{align} \label{eq:def-w}
W_i \equiv \sum_{k=1}^N \tilde{C}'_{ik}K_k
\end{align}
and observe that
\begin{align}
\sum_{k=1}^N K_k \tilde{C}'_{kj} = \sum_{k=1}^N \tilde{C}'^{*}_{jk} K_k = W_j^{*}.\label{cont}
\end{align}
Therefore,
\begin{align}
[(D_i + D_j)(k^2+p_n^2) + \mathrm{i} k_x (v_i - v_j)] \tilde{C}'_{ij} + K_i W_j^{*} + W_i K_j = 2 (k^2+p_n^2) \delta_{ij}.
\end{align}
This  can be solved for the matrix   \(\tilde{C}'_{ij}\) in terms of the vector $W_i$ to give
\begin{align} \label{eq:cij}
\tilde{C}'_{ij} = \frac{\delta_{ij}}{D_i} - \frac{K_i W_j^{*} + W_i K_j}{(D_i + D_j)(k^2+p_n^2) + \mathrm{i} k_x (v_i - v_j)}.
\end{align}
We can now write a closed equation for \(W_i\) by using the definition in Eq. (\ref{eq:def-w}) and  contracting the index \(j\) in Eq. (\ref{eq:cij}) with \(K_j\), yielding
\begin{align}
W_i = \frac{K_i}{D_i} - \sum_{j=1}^N \frac{K_i K_j W_j^{*} + W_i K_j^2}{(D_i + D_j)(k^2+p_n^2)+ \mathrm{i} k_x (v_i - v_j)}.
\end{align}
In vector form this then reads
\begin{align} \label{eq:W}
\mathbf{W} = \mathbf{S} - V \mathbf{W}^{*} - U \mathbf{W},
\end{align}
with
\begin{subequations}
\begin{align}
  S_i &\equiv \frac{K_i}{D_i}, \\
  V_{ij} &\equiv \frac{K_i K_j}{(D_i+D_j) (k^2+p_n^2) + \mathrm{i} k_x (v_i - v_j)}, \\
  U_{ij} &\equiv \delta_{ij} \sum_{k=1}^N \frac{K_k^2}{(D_i + D_k)(k^2+p_n^2) + \mathrm{i} k_x (v_i - v_k)}.
\end{align}
\end{subequations}
Taking the complex conjugate of Eq.~\eqref{eq:W} then gives
\begin{equation}
{\bf W}^* = {\bf S} - V^* {\bf W} - U^* {\bf W}^*,
\end{equation}
consequently
\begin{equation}
{\bf W}^*= (I+U^*)^{-1}({\bf S} - V^* {\bf W}),
\end{equation}
where $I$ denotes the identity matrix. Thus
\begin{equation}
{\bf W} = {\bf S} - V [(I+U^*)^{-1}({\bf S} - V^* {\bf W})] - U {\bf W}.
\end{equation}
Finally we obtain an equation solely for the vector ${\bf W}$ 
\begin{equation}
{\bf W}= [I+U -V (I+U^*)^{-1} V^*]^{-1}[I- V (1+U^*)^{-1}]{\bf S} .\label{eqW}
\end{equation}
This means that the computation of the correlation function has been reduced to a
problem of \(N \times N\) matrix inversion rather than the original Lyapunov form which is a problem of \(N^2(N+1)^2/4\) matrix inversion. The resulting problem is still rather complex but more palatable for computer algebra.
\section{Thermal van der Waals force for electrically driven ternary electrolytes}
\label{vdw}
We now apply the above reduction formula to the  problem of calculating the long-range van der Waals interaction  in driven electrolytes. In \cite{duCorrelationDecouplingCasimir2024,Du2025b} it was shown that the total  force (per unit area) is given by
\begin{equation}
f_{\text{t}} = f_{\rm vdW} + f_\text{ion},
\end{equation}
where $f_{\rm vdW}$ is given by Eq. (\ref{vdWE}) and $f_\text{ion}$ is the ionic contribution given by \cite{Du2025b}
\begin{align}
f_{\text{ion}} = \frac{2k_{\text{B}}T}{H}   \int \frac{\mathrm{d}\mathbf{k}}{(2\pi)^2} \sideset{}{'}\sum_{n=0}^{\infty} S_n(\mathbf{k}),\label{integral}
\end{align}
where the prime on the sum indicates that the term with $n=0$ is taken with a factor of $1/2$, and 
\begin{align} \label{eq:sn}
S_n({\bf k}) = \frac{1}{2} \left( N \frac{k^2}{k^2 + p_n^2} + \frac{p_n^2}{k^2 + p_n^2} \sum_{i=1}^N \frac{\tilde{C}_{ii}}{\bar{\rho}_i} \right),
\end{align}
where $|{\bf k}|=k$ denotes the wave number. We thus see that one needs to know the diagonal terms of the density-density correlation function to compute the force. 
The term necessary to compute $f_\text{ion}$ can then be written in terms of the vector $\bf W$ via $X$ defined as 
\begin{align} \label{eq:diagonal}
X \equiv \sum_{i=1}^N \frac{\tilde{C}_{ii}}{\bar{\rho}} = \sum_{i=1}^N D_i \tilde{C}'_{ii} = N - \sum_{i=1}^N \frac{K_i W'_i}{k^2 + p_n^2},
\end{align}
where \(W_i'= (W_i + W_i^{*})/2\) is the real part of \(W_i\) and we have used Eq.~(\ref{eq:cij}).

To proceed we split \(V\), \(U\) and \(W\) into their real and imaginary parts
\begin{align}
V = V' + \mathrm{i} V'', \quad
U = U' + \mathrm{i} U'', \quad
W = W' + \mathrm{i} W'',
\end{align}
each of which  can be explicitly written as
\begin{subequations}
\begin{align}
& V'_{ij} = \frac{K_i K_j (D_i + D_j) (k^2+p_n^2)}{(D_i + D_j)^2 (k^2 + p_n^2)^2 + k_x^2 (v_i - v_j)^2}, \\
& V''_{ij} = \frac{-K_i K_j k_x (v_i - v_j)}{(D_i + D_j)^2 (k^2+ p_n^2)^2 + k_x^2 (v_i - v_j)^2}, \\
& U'_{ij} = \delta_{ij} \sum_{k=1}^N \frac{K_k^2 (D_i + D_k) (k^2+p_n^2)}{(D_i + D_k)^2 (k^2+ p_n^2)^2 + k_x^2(v_i - v_k)^2}, \\
& U''_{ij} = \delta_{ij} \sum_{k=1}^N \frac{-K_k^2 k_x (v_i - v_k)}{(D_i + D_k)^2 (k^2+ p_n^2)^2 + k_x^2(v_i - v_k)^2}.
\end{align}
\end{subequations}
The real and imaginary parts of Eq. (\ref{eq:W})  are then given by
\begin{subequations}
\begin{align}
& (I + V' + U') W' + (V'' - U'') W'' = S, \\
& (V'' + U'') W' + (I - V' + U') W'' = 0.
\end{align}
\end{subequations}
Substiting the second equation into the first one leads to
\begin{align}
W' = [ (I + V' + U') - (V'' - U'')(I - V' + U')^{-1} (V'' + U'')]^{-1} S.
\end{align}
Clearly the above solution only requires the inversion of $N\times N$ matrices which is the key simplification.

We are interested in the long-range behavior of the van der Waals interaction.
To extract the long-range limit, we rescale
\begin{align}
k \to sk, \quad p_n \to s p_n, \quad k_x \to s k_x,
\end{align}
and we will consider the limit of small $s$ which governs the large $H$ behavior of the force. We now extract the term of the second order of \(s\) in \(W'\) denoted by $W'^{(2)}$, see Eqs.~(\ref{integral}--\ref{eq:diagonal}).
We make the following expansions
\begin{subequations}
\begin{align}
& V' = \sum_{n=-1}^\infty s^{2n} V'^{(2n)}, \\
& U' = \sum_{n=-1}^\infty s^{2n} U'^{(2n)}, \\
& V'' = \sum_{n=-1}^\infty s^{2n+1} V''^{(2n+1)}, \\
& U'' = \sum_{n=-1}^\infty s^{2n+1} U''^{(2n+1)}.
\end{align}
\end{subequations}
and list the first few terms that are relevant
\begin{subequations}
\begin{align}
& V'^{(-2)}_{ij} = U'^{(-2)}_{ij} = \delta_{ij} \frac{K_i^2}{2 D_i (k^2+p_n^2)}, \\
& V'^{(0)}_{ij} = 
\begin{cases}
\frac{K_i K_j (D_i + D_j) (k^2+p_n^2)}{k_x^2 (v_i - v_j)^2}, & i \neq j \\
0, & i = j
\end{cases}, \\
& U'^{(0)}_{ij} = \delta_{ij} \sum_{k=1,k\neq i}^N \frac{K_k^2 (D_i + D_k) (k^2+p_n^2)}{k_x^2 (v_i - v_k)^2}, \\
& V''^{(-1)}_{ij} = 
\begin{cases}
\frac{- K_i K_j }{k_x (v_i - v_j)}, & i \neq j \\
0, & i = j
\end{cases}, \\
& U''^{(-1)}_{ij} = \delta_{ij} \sum_{k=1,k\neq i}^N \frac{- K_k^2 }{k_x (v_i - v_k)},
\end{align}
\end{subequations}
where we have assumed \(v_i\neq v_j\) for $i\neq j$.
Note that in the case where $v_i=v_j$ for a given $i \neq j$ (which must therefore have the same charge), the problem 
actually simplifies but we do not consider this particular case in what follows. 
We then finally obtain
\begin{align} \label{eq:W-real-2}
W'^{(2)} = \left[ V'^{(-2)} + U'^{(-2)} - \left[ V''^{(-1)} - U''^{(-1)} \right] \left[ I - V'^{(0)} + U'^{(0)} \right]^{-1} \left[ V''^{(-1)} + U''^{(-1)} \right]  \right]^{-1} S.
\end{align}
We now extract the factors of \(k^2+p_n^2\) and \(k_x\)
from Eq.~\eqref{eq:W-real-2} and denote
\begin{subequations}
\begin{align}
& \tilde{V}'^{(-2)}_{ij} = \tilde{U}'^{(-2)}_{ij} = \delta_{ij} \frac{K_i^2}{2 D_i}, \\
& \tilde{V}'^{(0)}_{ij} = 
\begin{cases}
\frac{K_i K_j (D_i + D_j) }{(v_i - v_j)^2}, & i \neq j \\
0, & i = j
\end{cases}, \\
& \tilde{U}'^{(0)}_{ij} = \delta_{ij} \sum_{k=1,k\neq i}^N \frac{K_k^2 (D_i + D_k) }{ (v_i - v_k)^2}, \\
& \tilde{V}''^{(-1)}_{ij} = 
\begin{cases}
   \frac{- K_i K_j }{ (v_i - v_j)}, & i \neq j \\ 
   0, & i = j
\end{cases}, \\
& \tilde{U}''^{(-1)}_{ij} = \delta_{ij} \sum_{k=1, k\neq i}^N \frac{- K_k^2 }{ (v_i - v_k)}.
\end{align}
\end{subequations}
Then Eq.~\eqref{eq:W-real-2} can be written as 
\begin{align} \label{eq:W-real-2-var}
\frac{W'^{(2)}}{k^2+p_n^2} = \left[ \tilde{V}'^{(-2)} + \tilde{U}'^{(-2)} - \left[ \tilde{V}''^{(-1)} - \tilde{U}''^{(-1)} \right] \left[ \frac{k_x^2}{k^2+p_n^2} I - \tilde{V}'^{(0)} + \tilde{U}'^{(0)} \right]^{-1} \left[ \tilde{V}''^{(-1)} + \tilde{U}''^{(-1)} \right]  \right]^{-1} S.
\end{align}
For the ternary (three species) electrolyte case, the above reduction method leads to a  tractable solution for $W'^{(2)}$ using {\em Mathematica} and results in the long-range result
\begin{align}
X^{\text{lr}} = 3 - 1 -  \frac{k_x^2 \frac{\beta \epsilon E^2}{\sum_{i=1}^3 \bar{\rho}_i}}{k^2 + k_x^2 \frac{\beta \epsilon E^2}{\sum_{i=1}^3 \bar{\rho}_i}} - \frac{k_x^2 \frac{\beta \epsilon E^2}{\sum_{i=1}^3 \bar{\rho}_i}g}{ k^2 + k_x^2 \frac{ \beta\epsilon E^2}{\sum_{i=1}^3 \bar{\rho}_i}g},
\end{align}
where
\begin{align}
g =  \frac{\left(\sum_{i=1}^3 \bar{\rho}_i\right)\left[ \prod_{1\leq j < k \leq 3}^3 (D_j q_j - D_k q_k)^2  \right]}{\left( \sum_{i=1}^3 D_i q_i^2 \bar{\rho}_i \right)^2 \left\{\left[\sum_{1\leq j < k \leq 3}^3 \frac{D_j^2 D_k^2}{\bar{\rho}_j \bar{\rho}_k}(q_j^2 \bar{\rho}_j + q_k^2 \bar{\rho}_k) \right] - 2 \left(\prod_{j=1}^3 D_j q_j\right) \left( \sum_{k=1}^3 \frac{D_k}{\bar{\rho}_k q_k} \right)\right\}},\label{eqg}
\end{align}
which is a dimensionless variable.
Using this to compute the long-range part of the force via Eqs.~(\ref{integral}--\ref{eq:diagonal}) then gives
\begin{align}
f^{\text{lr}}_{\text{t}}(H) = \frac{\zeta(3)}{8 \pi \beta H^3} \left[
2 - \frac{1}{\sqrt{1 + \frac{\beta \epsilon E^2}{\sum_{i=1}^3 \bar{\rho}_i}}} - \frac{1}{\sqrt{1 +  g\frac{\beta \epsilon E^2 }{\sum_{i=1}^3\bar{\rho}_i}}} \label{flr}
 \right].
\end{align}
This result can be written in terms of an effective constant $\mathcal{H}(g, \Gamma)$, where $\Gamma = \frac{\beta \epsilon E^2 }{\sum_{i=1}^3\bar{\rho}_i}$ is proportional to  the ratio of the Maxwell pressure to the osmotic pressure. Putting this together thus gives
\begin{align}
f^{\text{lr}}_{\text{t}}(H) =  \frac{k_{\text{B}}T\zeta(3) }{8\pi H^{3}} \mathcal{H}(g,\Gamma),
\end{align}
where
\begin{align} \label{eq:h-constant}
\mathcal{H}(g, \Gamma) =
2 - \frac{1}{\sqrt{1 + \Gamma}} - \frac{1}{\sqrt{1 +  g \Gamma}}.
\end{align}
Eqs. (\ref{eqg}) and (\ref{eq:h-constant}) constitute the main results of this paper.

\section{Physical discussion of results}\label{disc}
We see that the expression for $g$ in Eq. (\ref{eqg})   does generically depend on the diffusion constants of the ions and consequently the third term on the right hand side for the force in Eq. (\ref{flr}) does also (the other  terms are however independent of the diffusion constants). We thus see that in general the result does depend on the diffusion constants showing conclusively that we are dealing  with a  nonequilibrium phenomenon. 

Another  fact to notice is that if there are two values of $i$ and $j$ such that $D_iq_i=D_jq_j$ then $g=0$. In this case the bare velocities of the two species due to the applied electric field $v_i= \beta D_i q_i E$ are the same.
One should notice that the case of a binary electrolyte can be recovered by setting $\bar{\rho}_3=0$, in this case we see that $g=0$ and we recover the result for a binary electrolyte given in Eq. (\ref{2elec}). When all the $D_i=D$ are equal, the result is independent of $D$ and we find
\begin{align}
g = \frac{\bar{\rho}_{1} \bar{\rho}_2 \bar{\rho}_3 (q_1 - q_2)^2 (q_2 - q_3)^2(q_{3}-q_1)^2}{(\sum_{i=1}^3 q_i^2 \bar{\rho}_i)^3 },
\end{align}
where we have used the electroneutrality condition Eq.~(\ref{eneut}).
This generalizes the results given in Ref.~\cite{Du2025b} for systems with uniform diffusivities.

If we consider the case where one of the species, say species 3, is much less mobile than the other two, so  $D_3\to 0$, we find
\begin{align}
g = \frac{(\sum_{i=1}^3 \bar{\rho}_i)q_1^2 q_2^2(D_1 q_1 - D_2 q_2)^2\bar{\rho}_1 \bar{\rho}_{2}}{(q_1^2 \bar{\rho}_1 + q_2^2 \bar{\rho}_2)(D_1 q_1^2 \bar{\rho}_1 + D_2 q_2^2 \bar{\rho}_2)^2}.
\end{align}
This result appears to be independent of the charge $q_3$, but there is in fact a dependence on $q_3$ via the electroneutrality constraint Eq. (\ref{eneut}).

In the case where species 3 is much more mobile than the other two species and hence the limit $D_3\to \infty$ (while keeping $\bar{\rho}_3$ nonzero) we find
\begin{align}
    g = \frac{(\sum_{i=1}^3 \bar{\rho}_i)(D_1 q_1 - D_2 q_2)^2 \bar{\rho}_1 \bar{\rho}_2}{\bar{\rho}_3[(D_1^2 \bar{\rho}_2 + D_2^2 \bar{\rho}_1)(\sum_{i=1}^3 q_i^2 \bar{\rho}_i) -(D_1 q_2\bar{\rho}_2 + D_2 q_1 \bar{\rho}_1)^2]}.
\end{align}

\begin{figure}[ht]
  \centering
  \includegraphics[width=0.9\linewidth]{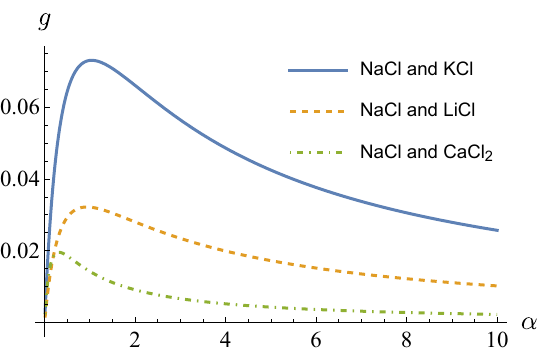}
  \caption{\label{fig:g-alpha} From top to bottom, $g$ as a function of $\alpha = \bar{\rho}_2 /\bar{\rho}_1$ for solutions of NaCl/KCl, NaCl/LiCl and NaCl/CaCl$_2$ computed from Eq. (\ref{eqg}).}
\end{figure}

As a concrete example let us consider an electrolyte solution composed of NaCl with density $\bar{\rho}_{\text{NaCl}}$ and KCl with density $\bar{\rho}_{\text{KCl}}$. If we define by species $1$ Na$^+$, species 2 K$^+$ and species 3
Cl$^-$, this means that $q_1 = q_2 = - q_3=e$, where $e$ is the charge of the electron, and assuming that they are strong (fully disassociated electrolytes) we have $\bar{\rho}_1 = \bar{\rho}_{\text{NaCl}}$, $\bar{\rho}_2 = \bar{\rho}_{\text{KCl}}$ and $\bar{\rho}_3 = \bar{\rho}_{\text{NaCl}} + \bar{\rho}_{\text{KCl}}$. For the simple model used here, there is no difference in the static electrical properties of Na$^+$ and K$^+$ and so, other than being distinguishable, the two salts are equivalent. However the respective diffusion constants are given by $D_{\text{K}^+}=1.96\times 10^{-9} m^2s^{-1}$, $D_{\text{Na}^+}=1.33\times 10^{-9} m^2s^{-1}$ and $D_{\text{Cl}^-}=2.03\times 10^{-9} m^2s^{-1}$~\cite{TableDiffusionCoefficients}. One can also consider the case of a mixture of NaCl and LiCl, where $D_{\text{Li}^+}=1.03\times 10^{-9} m^2s^{-1}$ (so here $\bar{\rho}_2 = \bar{\rho}_{\text{LiCl}}$). We can then plot $g$ as a function of $\alpha = \bar{\rho}_2 /\bar{\rho}_1$. Fig. (\ref{fig:g-alpha}) shows this for NaCl and KCl mixtures (solid curve) and NaCl and LiCl mixtures (dashed curve). Notice that in both cases $g\to0$ as $\alpha\to0$ and $\alpha\to \infty$ which corresponds to a single salt type in solution (respectively pure NaCl or pure KCl or LiCl) because we are effectively in the binary electrolyte case.
Looking at Fig.~(\ref{fig:g-alpha}), we see that $g$ is very small in the examples given.
Notice that the nonequilibrium effect is stronger for the case of KCl,  essentially because the difference between the diffusion constants of K$^+$ and
Na$^+$ is larger than that between Li$^+$ and Na$^+$. Also shown in Fig. (\ref{fig:g-alpha}) is the curve for mixtures of NaCl and CaCl$_{2}$ (dot-dashed curve), so in this case $\bar{\rho}_1= \bar{\rho}_{\text{NaCl}}$, $\bar{\rho}_2= \bar{\rho}_{\text{CaCl}_2}$ and ${\bar\rho}_3=\bar{\rho}_{\text{NaCl}}+ 2 \bar{\rho}_{\text{CaCl}_2}$ and as well $q_2= 2q_1$ and $q_1=-q_3=e$. We also use that $D_{\text{Ca}^{2+}}= 0.793\times 10^{-9} m^2s^{-1}$.
Interestingly the effect here is smaller than the other two cases, despite the difference in the charge of the cations. The reason for this is that  $2e D_{\text{Ca}^{2+}}$ is close to $eD_{\text{Na}^{+}}$ and the two cationic types have similar bare velocities, thus reducing the nonequilibrium effect.

\begin{figure}[ht]
  \centering
  \includegraphics[width=0.9\linewidth]{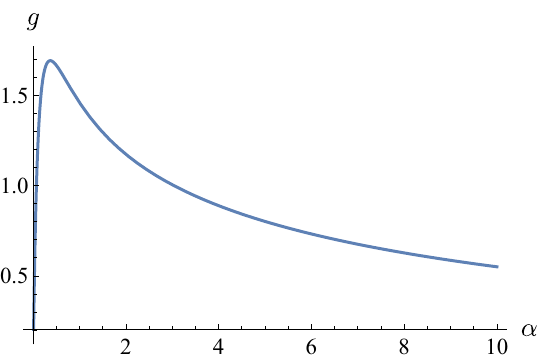}
  \caption{\label{fig:g-alpha_LiCl-HCl} $g$ as a function of $\alpha = \bar{\rho}_2 /\bar{\rho}_1$ for solution of LiCl/HCl computed from Eq. (\ref{eqg}).}
\end{figure}

The dominant factor in the expression for $g$ which determines its size is the factor $(D_{1}q_{1} - D_{2} q_{2})^{2}$, where we assume $q_{1}$ and $q_{2}$ have the same sign.
In order to maximize this term, one must maximize $D_{2}q_{2}$ while minimizing $D_{1}q_{1}$. One way to do this is to take the species 2 to be H$^{+}$, which has $D_{\text{H}^{+}} = 9.31 \times 10^{-9} m^2s^{-1}$, while taking the species 1 to be Li$^{+}$, which has a relatively small diffusion constant, and again taking the anion to be Cl$^{^{-}}$.
Shown in Fig.~(\ref{fig:g-alpha_LiCl-HCl}) is $g$ as a function of $\alpha = \bar{\rho}_{\text{HCl}}/\bar{\rho}_{\text{LiCl}}$. Here the effect is much larger and we see $g$ can be of order 1.

\begin{figure}[ht]
  \centering
  \begin{subfigure}[b]{.5\textwidth}
    \centering
    \includegraphics[width=.9\linewidth]{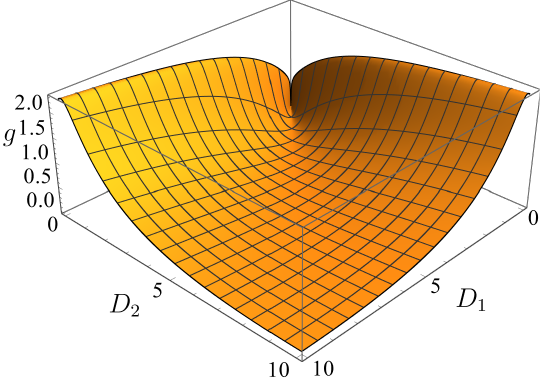}
    \caption{}
  \end{subfigure}
  \begin{subfigure}[b]{.5\textwidth}
    \centering
    \includegraphics[width=.9\linewidth]{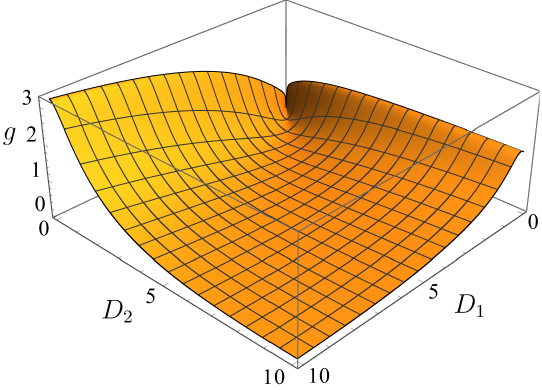}
    \caption{}
  \end{subfigure}%
  \caption{\label{2dplotsNaClKCl}$g$ as a function of $D_1$ and $D_2$ in units where $D_3 = 1$ for an NaCl/KCl type system. (a) $\bar{\rho}_{\text{NaCl}}= \bar{\rho}_{\text{KCl}}$ . (b) $\bar{\rho}_{\text{KCl}}= 2\bar{\rho}_{\text{NaCl}}$.}
\end{figure}

In Fig. (\ref{2dplotsNaClKCl}) we show how $g$ varies at fixed charges and densities as $D_1$ and $D_2$ (say the cationic diffusion constants) are varied in units where  (say anionic diffusion constant) $D_3=1$  relevant to the NaCl/KCl type system. In  Fig. (\ref{2dplotsNaClKCl}a) we show the case where $\bar{\rho}_{\text{NaCl}}= \bar{\rho}_{\text{KCl}}$. We see that  $g$ is symmetric about the line $D_1=D_2$ where it vanishes, and we see that the maximal effect is obtained when $D_1$ and $D_2$ are very different (one large and the other small). In Fig. (\ref{2dplotsNaClKCl}b) we show the case for a NaCl/KCl type system when $\bar{\rho}_{\text{KCl}}=2 \bar{\rho}_{\text{NaCl}}$, here the symmetry is broken but $g$ of course still vanishes when $D_1=D_2$, the maximal effect is achieved when $D_2$ is small and $D_1$ is large, so the species with the largest concentration has the smaller diffusivity.

\begin{figure}[ht]
  \centering
  \begin{subfigure}[b]{.5\textwidth}
    \centering
    \includegraphics[width=\linewidth]{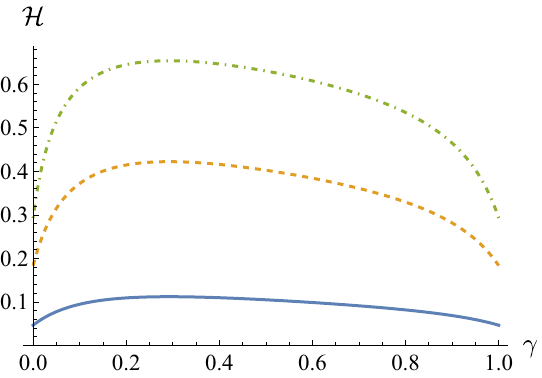}
    \caption{}
  \end{subfigure}
  \begin{subfigure}[b]{.5\textwidth}
    \centering
    \includegraphics[width=\linewidth]{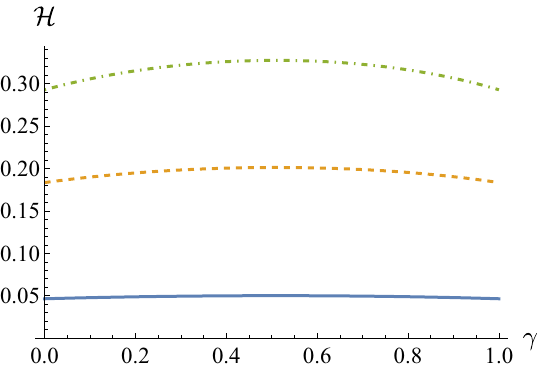}
    \caption{}
  \end{subfigure}
  \caption{\label{fig:h-gamma} $\mathcal{H}$ as a function of $\gamma$, where $\Gamma=0.1$ (solid curves), $\Gamma=0.5$ (dashed curves) and $\Gamma = 1$ (dot-dashed curves). (a) NaCl and HCl. (b) NaCl and KCl.}
\end{figure}

We now discuss the modification of thermal van der Waals force via its dependence on the effective constant $\mathcal{H}(g, \Gamma)$.
As we are interested in the dynamical dependence of the force, we consider a situation where the two monovalent salts have the same anion but differing cations of the same charge $q$ but different diffusion constants $D_{1}$ and $D_{2}$.
We take the concentration of salt 1 to be $(1-\gamma)\rho$ and the concentration of salt 2 to be $\gamma \rho$. Electroneutrality means that $\bar{\rho}_{1} = (1-\gamma)\rho$, $\bar{\rho}_{2} = \gamma \rho$ and $\bar{\rho}_{3} = \rho$.
In terms of its static electrical properties the solution is therefore independent of $\gamma$.
This gives $\Gamma = \frac{\beta \epsilon E^{2}}{2\rho}$, which is also independent of $\gamma$. Therefore, the only dependence on $\gamma$ is through $g$ due to the difference between $D_{1}$ and $D_{2}$.
Let us consider a case where a large value of $g$ can be obtained, for example, where salt 1 is NaCl and salt 2 is HCl.
We plot in Fig.~(\ref{fig:h-gamma}) $\mathcal{H}(\Gamma,\gamma)$ as a function of $\gamma$ (because $g$ is independent of $\rho$) for various values of $\Gamma$, which can be tuned independently.
In Fig.~(\ref{fig:h-gamma}a), we see of course that the effect is much larger as $\Gamma$ increases from $0.1$ to $1$. In~\cite{duCorrelationDecouplingCasimir2024,Du2025b}, it was pointed out that values of $\Gamma$ up to $0.5$ may be feasible while avoiding electrolysis.
The maximal effect is observed for $\gamma$ around $0.25$ and minimal values are given for $\gamma=0$ and $\gamma=1$, which corresponds to binary electrolytes.
In Fig.~(\ref{fig:h-gamma}b), the effect is much smaller due to the fact that Na$^{+}$ and K$^{+}$ have similar diffusion constants. In these cases, the maximal effect occurs at a larger value $\gamma\approx 0.5$.

\section{Conclusion}
\label{conc}
The application of SDFT in the linearized approximation has emerged as a powerful method to study the physics of electrolytes out of equilibrium. Despite the fact that the theory becomes Gaussian in this approximation, the physics emerging is encoded in Lyapunov equations which are linear but with a complexity which increases with the number of electrolyte species. This suggests that multispecies electrolyte solutions could exhibit new and interesting physics when driven out of equilibrium compared to binary electrolytes. In this paper we have presented a considerable algebraic simplification to the analysis of these Lyapunov equations which allows the linear system size to be reduced from $N(N+1)/2$ to $N$. Using this simplification we have been able to obtain closed-form expressions for the long-range force induced between two dielectric surfaces when the intervening electrolyte is driven by an electric field applied parallel to the surfaces. It is somewhat surprising, given the out of equilibrium nature of the problem, that the force for a binary electrolyte is independent of the diffusion constants of the ionic species. Here, by studying the case of ternary electrolytes, we see that the result for binary electrolytes is somewhat fortuitous and that for ternary electrolytes there is a dependence on the ionic diffusion constants.
Our analysis has shown that the nonequilibrium effect as quantified by the value of $g$ is maximized when the two species 1 and 2 of the same charge sign in the electrolyte have very different values of $D_1 q_1$ and $D_2q_2$.
Maximizing this difference by varying the charge $q_{1}$ with respect to $q_{2}$ (for example, choosing $q_{2} = 2q_{1}$) tends to lead to relatively small differences as cations of valency 2 typically have diffusion constants which are close to $50\%$ of the diffusion constants of cations of valency 1.
However, if one takes H$^{+}$  as one of the cations, its large diffusion constant relative to other cations of the same charge can lead to large values of $g$.
In general this leads us to suspect that novel nonequilibrium effects in electrolyte solutions may be more likely when one of the electrolyte species is an acid, i.e., contains H$^+$ cations.

Further extensions of this work include generalization to systems with surface charges, however, one must develop a version of SDFT that can deal with inhomogeneities. This presents a significant technical challenge. Another course of study will be to analyze these systems when AC fields are applied, this is particularly important as for AC fields, the experimental protocols are more established \cite{richterIonsACElectric2020}.

\begin{acknowledgments}
G.D.~and  B.M.~acknowledge funding from the Key Project No.~12034019 of the National Natural Science Foundation of China. D.S.D.~acknowledges support  from the grant No.~ANR-23-CE30-0020 EDIPS, and by the European Union through the European Research Council by the EMet-Brown (ERC-CoG-101039103) grant.
\end{acknowledgments}

\bibliography{ms.bib}
\end{document}